\documentstyle[preprint,aps]{revtex}

\newcommand{\str}        {$\rm^{1}$}
\newcommand{\gsi}        {$\rm^{2}$}
\newcommand{\legnaro}    {$\rm^{3}$}
\newcommand{\clt}        {$\rm^{4}$}
\newcommand{\bucarest}   {$\rm^{5}$}
\newcommand{\zagreb}     {$\rm^{6}$}
\newcommand{\itep}       {$\rm^{7}$}
\newcommand{\dresde}     {$\rm^{8}$}
\newcommand{\budapest}   {$\rm^{9}$}
\newcommand{\warsaw}     {$\rm^{10}$}
\newcommand{\rrik}       {$\rm^{11}$}
\newcommand{\heidelberg} {$\rm^{12}$}

\begin{document}
\draft

\title{Onset of Nuclear Matter Expansion in Au+Au Collisions}

\author{
P.~Crochet\str\footnote{Present address : GSI, Darmstadt, Germany}, 
F.~Rami\str,
A.~Gobbi\gsi,
R.~Don\`a\str$^{,}$\legnaro,
J.P.~Coffin\str,
P.~Fintz\str,
G.~Guillaume\str,
F.~Jundt\str,
C.~Kuhn\str,
C.~Roy\str,
B.~de~Schauenburg\str,
L.~Tizniti\str,
P.~Wagner\str,
J.P.~Alard\clt,
V.~Amouroux\clt,
A.~Andronic\bucarest,
Z.~Basrak\zagreb,
N.~Bastid\clt,
I.~Belyaev\itep,
D.~Best\gsi,
J.~Biegansky\dresde,
A.~Buta\bucarest,
R.~\v{C}aplar\zagreb,
N.~Cindro\zagreb,
P.~Dupieux\clt,
M.~D\v{z}elalija\zagreb,
Z.G.~Fan\clt,
Z.~Fodor\budapest,
L.~Fraysse\clt,
R.P.~Freifelder\gsi,
N.~Herrmann\gsi,
K.D.~Hildenbrand\gsi,
B.~Hong\gsi,
S.C.~Jeong\gsi,
J.~Kecskemeti\budapest,
M.~Kirejczyk\warsaw,
P.~Koncz\budapest,
M.~Korolija\zagreb,
R.~Kotte\dresde,
A.~Lebedev\itep$^{,}$\rrik,
Y.~Leifels\gsi,
V.~Manko\rrik,
D.~Moisa\bucarest,
J.~M\"osner\dresde,
W.~Neubert\dresde,
D.~Pelte\heidelberg,
M.~Petrovici\bucarest,
C.~Pinkenburg\gsi,
P.~Pras\clt,
V.~Ramillien\clt,
W.~Reisdorf\gsi,
J.L.~Ritman\gsi,
A.G.~Sadchikov\rrik,
D.~Sch\"ull\gsi,
Z.~Seres\budapest,
B.~Sikora\warsaw,
V.~Simion\bucarest,
K.~Siwek-Wilczy\'nska\warsaw,
U.~Sodan\gsi,
K.M.~Teh\gsi,
M.~Trzaska\heidelberg,
M.~Vasiliev\rrik,
G.S.~Wang\gsi,
J.P.~Wessels\gsi,
T.~Wienold\gsi,
K.~Wisniewski\gsi,
D.~Wohlfarth\dresde,
A.~Zhilin\itep \\
(FOPI Collaboration)
}

\address{
\str  Centre de Recherches Nucl\'eaires, IN2P3-CNRS, Universit\'e
Louis Pasteur, Strasbourg, France \\
\gsi Gesellschaft f\"ur Schwerionenforschung, Darmstadt, Germany \\
\legnaro Istituto Nazionale di Fisica Nucleare, Legnaro, Italy \\
\clt Laboratoire de Physique Corpusculaire, IN2P3-CNRS,
Universit\'e Blaise Pascal, Clermont-Ferrand, France\\
\bucarest Institute for Physics and Nuclear Engineering, Bucharest, Romania \\
\zagreb Ru{d\llap{\raise 1.22ex\hbox
{\vrule height 0.09ex width 0.2em}}\rlap{\raise 1.22ex\hbox
{\vrule height 0.09ex width 0.06em}}}er
Bo\v{s}kovi\'{c} Institute, Zagreb, Croatia \\
\itep Institute for Theoretical and Experimental Physics, Moscow, Russia \\
\dresde Forschungszentrum Rossendorf, Dresden, Germany \\
\budapest Research Institute for Particles and Nuclear Physics, 
Budapest, Hungary \\ 
\warsaw Institute of Experimental Physics, Warsaw University, Warsaw, Poland \\
\rrik Russian Research Institute ``Kurchatov", Moscow, Russia \\
\heidelberg Physikalisches Institut der Universit\"at Heidelberg,
Heidelberg, Germany \\
}

\maketitle

\begin{abstract}
Using the FOPI detector at GSI Darmstadt, excitation functions of 
collective flow components were measured for the Au+Au system, 
in the reaction plane and out of this plane,
at seven incident energies ranging from 100$A$\,MeV to 800$A$\,MeV.
The threshold energies, corresponding to the onset of sideward-flow
(balance energy) and squeeze-out effect (transition energy), are 
extracted from extrapolations of these excitation functions toward lower
beam energies for charged products with $Z \geq 2$.
The transition energy is found to be larger than the balance energy.
The impact parameter dependence of both balance and transition energies, 
when extrapolated to central collisions,
suggests comparable although slightly higher values than the threshold energy 
for the radial flow.
The relevant parameter seems to be the energy deposited into the system
in order to overcome the attractive nuclear forces. \\
\end{abstract}

\vspace{0.5cm}


\vspace{1.cm}


\section{Introduction}

Collective motions of nuclear matter occurring in heavy ion collisions
are of great interest since they are expected to provide information
about the properties of hot and dense nuclear matter and the underlying
equation of state (EoS)~\cite{sto 80}.
Flow effects were predicted by hydrodynamical 
calculations~\cite{sto 80,sch 74} and experimentally evidenced at 
LBL-BEVALAC~\cite{gut 89}.
At beam energies $E_{\rm lab} \geq 200$$A$\,MeV, the interaction between nuclei
is dominated by individual nucleon-nucleon scattering and the repulsive 
component of the mean field.
This leads to a collective deflection of matter to positive angles in the
reaction plane {\em i.e.} in the direction of the projectile remnants
(sideward-flow).
Conversely, at few tens of $A$\,MeV, the interaction is dominated 
by the attractive mean field, so that nucleons emitted in the reaction
plane are deflected to negative angles~\cite{tsa 86}.
At a certain intermediate incident energy, named the balance energy 
$E_{\rm BAL}$, 
the attractive component and the repulsive component of the interactions 
balance each 
other and consequently the flow crosses zero, changing from a negative sign 
at low energies to a positive sign at high energies.
The balance effect was extensively investigated at 
GANIL~\cite{sul 90,pet 92,she 93} and MSU~\cite{kro 89,ogi 90,wes 93,pak 96}
by measuring different colliding systems.
It was also studied in the framework of theoretical models
\cite{mol 85,xu 91,xu 92,mot 92,har 92,xu 93,kla 93,li 93,zho 93,ma 93,zho 95,sof 95,leh 96}.
For semi-central collisions, the balance energy was found to be 
sensitive to the stiffness of the nuclear EoS and to the in-medium 
reduction of the nucleon-nucleon cross section~\cite{xu 91}.
At lower impact parameters the balance energy is expected to be only
sensitive to the in-medium nucleon-nucleon cross section~\cite{xu 91,zho 93}.
Recent experimental results~\cite{pop 94,wil 95,but 95a,tsa 96}
revealed also, in the same beam energy range, 
signatures of a change in the azimuthal emission pattern of mid-rapidity 
particles, from an in-plane enhancement at low incident energies to the well 
known
out-of-plane preferential emission, the so-called 
squeeze-out~\cite{gut 89b,dem 90}, at 
higher energies.
The incident energy where this transition takes place (termed $E_{\rm TRA}$)
is also found to be sensitive to the in-medium nucleon-nucleon cross 
section~\cite{tsa 96}.
 
The excitation functions of both sideward-flow and squeeze-out
effects were measured with the
FOPI detector~\cite{gob 93} for the Au+Au system at seven incident energies
between $E_{\rm lab}=100$$A$\,MeV and $E_{\rm lab}=800$$A$\,MeV.
The ability of the FOPI device to detect intermediate mass fragments 
($Z \geq 3$) allows a cleaner identification of the collective flow signal
and a better extrapolation of the measured excitation functions
of the in-plane and of the out-of-plane flows toward their thresholds
at low incident energies. 
We found however that the threshold energies are rather insensitive to the 
charge of the considered particles ($Z > 1$).
We found also that the transition point seems to be located at higher beam 
energies than the balance point.
On the other hand, the FOPI detector ensures a wide range of 
impact parameter collisions to be explored.
This offers the possibility to investigate the centrality dependence
of $E_{\rm BAL}$ and $E_{\rm TRA}$, an aspect which was recognized
to be crucial in this kind of study~\cite{pak 96,xu 91,sof 95}.
This allows us to discuss for the first time 
this centrality dependence in conjunction
with the threshold energy ($E_{\rm RAD}$) 
for the radial flow in highly central collisions.
It shows that the three threshold energies ($E_{\rm BAL}$, $E_{\rm TRA}$, and
$E_{\rm RAD}$) might be attributed
to a common phenomenon, the relevant parameter being the energy 
deposited into the nuclear system.

\section{Experimental setup}

The data presented in this paper concern the Au+Au system at seven incident
energies $E_{\rm lab}=100$, 120, 150, 250, 400, 600 and 800$A$\,MeV. 
They have been collected with the Phase I of
the FOPI detector~\cite{gob 93} at the SIS/ESR accelerator facility, GSI 
Darmstadt.
In its Phase I configuration, the FOPI detector covers in full azimuth
the laboratory polar angles ($\Theta_{\rm lab}$) from $1.2^{\circ}$ to
$30^{\circ}$.
It consists mainly of a highly segmented Forward Wall of plastic 
scintillators divided into two parts : the Inner Wall made of 252
trapezoidal scintillators which covers the $\Theta_{\rm lab}$ range
between $1.2^{\circ}$ and $7.5^{\circ}$, and the Outer Wall made of 
512 scintillator strips which covers the $\Theta_{\rm lab}$ 
domain from $7^{\circ}$ to $30^{\circ}$.
The Forward Wall provides an element identification and the velocity
of the reaction products through energy loss and 
time-of-flight measurements.
Its segment structure allows to determine the velocity vector components.
A complementary shell of 188 thin energy loss detectors is mounted
in front of the Forward Wall in order to achieve lower detection 
thresholds.
This cluster detector is made of an ensemble of gas-filled ionisation
chambers (Parabola) mounted in front of the Outer Wall, and
thin plastic scintillator paddles (Rosace) combined with the Inner Wall.
In order to reduce the background scattering in the air gas,
a helium bag is placed
between the target and the detectors.
This setup measures simultaneously most of the light charged particles 
and intermediate mass fragments (up to $Z=15$) emitted in the forward 
center-of-mass (c.m.) hemisphere.
Its high granularity allows high multiplicity events to be measured
with a negligible multi-hit rate.
The apparatus ensures a very good azimuthal symmetry which is an important
feature for the study of the flow phenomena.

\section{Event characterisation}

\subsection{Impact parameter determination}

The measured events were sorted according to their degree of centrality
using the standard method based on the correlation between the
multiplicity of emitted particles and the impact parameter.
The multiplicity distribution of the charged particles detected in the
Outer Wall exhibits the typical plateau for intermediate values,
followed by a steep decrease at the highest multiplicities~\cite{ala 92}.
The highest multiplicity bin (named PM5) has been defined by cutting
at half of the plateau value 
(the corresponding lower limits of the PM5 multiplicity bin are given
in Tab.~\ref{tab1}).
The remaining part of the multiplicity 
distribution has been subdivided into four equally
spaced bins (named PM1 to PM4) according to the procedure used by the Plastic
Ball Collaboration~\cite{dos 85}.
The results presented in what follows include only events belonging
to the PM3-PM5 multiplicity classes where background contamination,
estimated from measurements without target, is negligible.
The mean impact parameter associated to each PM event class has been 
determined in the framework of the IQMD (Isospin Quantum Molecular
Dynamics) model~\cite{har 92,har 94} by filtering the theoretical 
calculations with the realistic simulator of the FOPI detector.
The resulting $<b>$ values and their r.m.s deviations 
are presented in Tab.~\ref{tab2} for the beam energies 
$E_{\rm lab}=150$ and 400$A$\,MeV. 
Note that $<b>$ values are the same within the r.m.s deviations 
at the other beam energies. 
It can be seen from Tab.~\ref{tab2} that the multiplicity 
criterium offers a large coverage of the impact parameter range.
On the other hand, model studies~\cite{cro 96a}, using the so-called 
quality factor introduced by Cugnon and L'H\^ote~\cite{cug 83}, 
allowed us to show that with the FOPI/Phase-I setup, the multiplicity
criterium, as compared to other criteria, appears as the most 
appropriate one for exploring flow observables over a large 
impact parameter domain.

\subsection{Reaction plane reconstruction}

The reaction plane was reconstructed with the transverse momentum
analysis devised by Danielewicz and Odyniec~\cite{dan 85}.
In order to remove autocorrelation effects, the azimuth of the reaction plane
was estimated for each particle ${\rm i}$ in a given event 
as the plane containing the vector $\vec{Q_{\rm i}}$ and the beam axis, where
$\vec{Q_{\rm i}}$ is calculated 
from the transverse momenta $\vec{p_{\rm t}^{j}}$ of all detected particles 
except the particle i : 
$$
\vec{Q_{\rm i}}\;=\;\sum_{{\rm j=1\atop j\neq i}}^{{\rm M}}\,\omega^{{\rm j}} 
(\vec{p_{\rm t}^{\rm j}} + m^{\rm j} {\vec{v_{\rm b}^{\rm i}}}).
$$
${\rm M}$ is the multiplicity of the event
and $\omega^{\rm j} = 1\;\mbox{if}\;y^{(0)}_{\rm j}>\delta,
-1\;\mbox{if}\;y^{(0)}_{\rm j}<-\delta$ and $0$ otherwise.
$y^{(0)}_{\rm j}$ is the ${\rm j}^{\rm th}$ particle rapidity 
divided by the projectile
rapidity in the c.m. system.
The parameter $\delta$, choosen equal to 0.5, was
introduced in order
to remove mid-rapidity particles which have a negligible correlation with the
reaction plane.
According to~\cite{ogi 89}, a boost velocity 
$\vec{v_{\rm b}^{\rm i}} = \vec{p_{\rm t}^{\rm i}}/(m^{\rm sys} -
m^{\rm i}) $ 
($m^{\rm i}$ is the mass of the particle ${\rm i}$ and $m^{\rm sys}$ 
is the sum of the projectile and target masses)
was applied to each particle ${\rm j}$  
in order to take into account the effects of momentum conservation due
to the exclusion of the particle ${\rm i}$.   
The influence of these effects on the observables considered in the 
present work was found to be of a few percent at the lowest beam
energies ($E_{\rm lab}=100$ and 120$A$\,MeV) and negligible at higher 
incident energies.

The accuracy of the reaction plane reconstruction ({\em i.e.}
the azimuthal deviation $\Delta\Phi_{\rm R}$ of the reconstructed reaction 
plane with respect to the true one) was estimated 
for each event class by randomly dividing each event into two equal 
parts and by taking the one half
of the angle between the $\vec{Q}$ vectors of the two 
subevents~\cite{dan 85}.
The azimuthal dispersion $\sigma(\Delta\Phi_{\rm R})$ was found to vary 
typically 
from $\sim 20^{\circ}$ to $\sim 40^{\circ}$ for the PM event classes 
under consideration.
It is worth noting that the use of a heavy system such as Au+Au offers,
in the beam energy range considered here,
a good event characterisation both in centrality and reaction plane 
reconstruction as compared to lighter systems whose 
measured ejectile multiplicities are lower.

\section{Results and discussion}

\subsection{Sideward-flow}

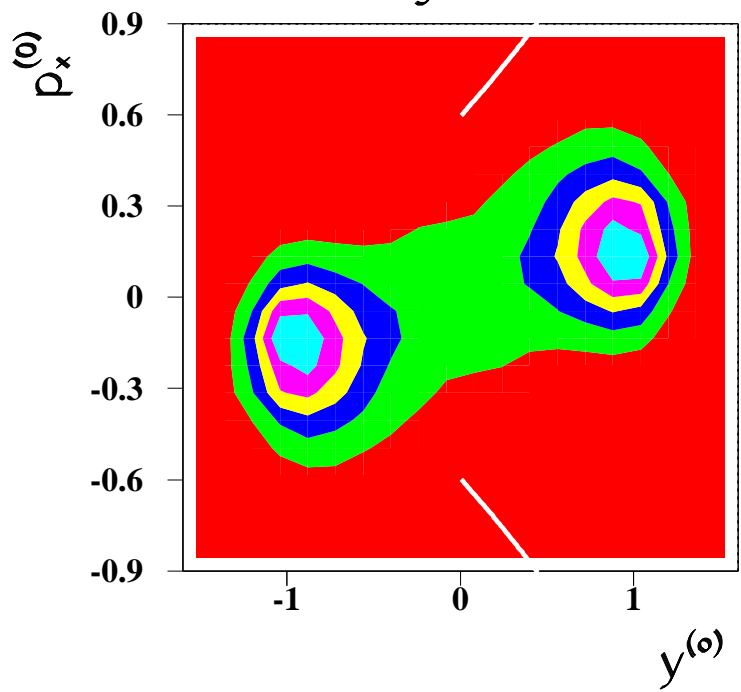
\begin{figure}
\caption{Upper panel : Normalized in-plane transverse momentum 
$p_{\rm x}^{(0)}$ {\em versus} the
normalized c.m. rapidity $y^{(0)}$ for $Z=4$ particles detected in
semi-central (PM4) Au+Au collisions at $E_{\rm lab}=250$$A$\,MeV.
The plot is obtained by assuming a forward/backward symmetry.
The different grey levels correspond to different linear cuts in multiplicity.
The $\Theta_{\rm c.m.}=30^{\circ}$ cut is represented by the solid white
curves. \\
Lower panel : Mean normalized in-plane transverse momentum $<p_{\rm x}^{(0)}>$
as a function of $y^{(0)}$ for $Z=4$ particles detected in semi-central (PM4)
Au(250$A$\,MeV)+Au collisions.
The data (open stars) are measured only for $y^{(0)} > 0$ and then
reflected around the origin (full stars).
Data points are larger than the corresponding statistical uncertainties.
The solid curve is the result of the fit described in the text.}
\label{fig1}
\end{figure}

The in-plane flow component (sideward-flow) is examined in terms of the
normalized in-plane transverse momentum $p_{\rm x}^{(0)}$ 
($p_{\rm x}^{(0)}=(p_{\rm x}/A)/p_{\rm c.m.}^{\rm p}$ 
where $p_{\rm x}/A$ is the in-plane transverse
momentum per nucleon and $p_{\rm c.m.}^{\rm p}$ is the projectile momentum per
nucleon in the c.m. system) as a function 
of the normalized c.m. rapidity $y^{(0)}$ ($y^{(0)}$ is defined above).
This normalisation, suggested in earlier works~\cite{bal 84,bon 87}, 
is motivated by the fact that one obtains a scale invariant
representation of the data
in a fluid dynamical description of the collision.
Figure~\ref{fig1} shows a typical example of $p_{\rm x}^{(0)}$ 
(Fig.~\ref{fig1}.a) and 
$<p_{\rm x}^{(0)}>$ (Fig.~\ref{fig1}.b) {\em versus} $y^{(0)}$
plots for $Z=4$ particles detected in semi-central (PM4) reactions
at $E_{\rm lab}=250$$A$\,MeV.
Since the experimental apparatus covers only the forward c.m. hemisphere, 
the plot has been measured for positive $y^{(0)}$ rapidities and reflected 
for negative ones.
As shown by Fig.~\ref{fig1}.b, the dependence of 
$<p_{\rm x}^{(0)}>$ on $y^{(0)}$
exhibits the well known S-shape behavior~\cite{dan 85}
demonstrating the collective transfer of momentum between the backward
and the forward hemispheres.
The linear part of the curve in the participant region ({\em i.e.}, 
at mid-rapidity) reflects the so-called side-splash effect while the 
fall-off starting just below the projectile rapidity ($y^{(0)}=1$)
is caused by the bounce-off effect~\cite{gus 84}.
A quantitative measure of the amount of flow in the participant 
region of the collision is given by the so-called normalized flow parameter
$F_{\rm S}^{(0)}$ which is commonly defined as the slope of the 
$<p_{\rm x}^{(0)}>$ {\em versus} $y^{(0)}$ curve at mid-rapidity~\cite{dos 86} :
$F_{\rm S}^{(0)}=\left. {\rm d} <p_{\rm x}^{(0)}> / 
{\rm d} y^{(0)}\right|_{y^{(0)} \simeq 0}$.
Technically the $F_{\rm S}^{(0)}$ parameter is obtained by fitting a polynomial
function of the form : $a + F_{\rm S}^{(0)}\times y^{(0)} + 
c\times (y^{(0)})^{3}$ 
to the data (Fig.~\ref{fig1}.b).
The fit was restricted to the linear branch of the S-shape curve.
As shown in reference~\cite{dan 85}, because the particle 
momenta are not projected onto the true reaction plane, 
their projections are on average biased downward by a factor 
$1/<\cos(\Delta\Phi_{\rm R})>$ where $\Delta\Phi_{\rm R}$, 
as mentioned before, is the estimate of the 
azimuthal deviation of the reconstructed reaction plane with respect to 
the true one.
The data shown in Fig.~\ref{fig1}.b and all the $F_{\rm S}^{(0)}$ 
values presented in 
what follows are corrected for this effect.
The correction factors ($1/<\cos(\Delta\Phi_{\rm R})>$) were typically 
ranging from 
1.10 to 1.45 depending on the beam energy and the multiplicity bin.

The precise evaluation of the acceptance effects on the in-plane flow
is rather difficult due to the complexity of the different experimental 
constraints.
This can only be investigated in the framework of realistic simulations
where theoretical calculations are passed through the detector filter.
In this context, we have used the IQMD model~\cite{har 92,har 94} 
which is known to reproduce quite well experimental flow 
data~\cite{ram 93,ramil 95,wie 93,cro 96c,don 96,rei 96}.
A few thousand of IQMD events were generated over a large range of impact
parameters with the HM choice (Hard EoS plus a momentum 
dependent potential) of the nuclear interaction.
This force is recognized as providing 
the best description of the observed trends in the in-plane flow 
data~\cite{ramil 95,wie 93,cro 96c,don 96}.
For the present study, theoretical events were filtered applying 
geometrical cuts and energy thresholds of the FOPI detector.
They were presorted in accordance with the above mentioned
procedure used for the data (see before).
The in-plane flow was extracted with respect to the true reaction plane which
is known in the model.
Because of the limited statistics, apparatus effects could only 
be evaluated for light particles.
We found that the experimental cuts only slightly affect
$<p_{\rm x}^{(0)}>$ values in the forward c.m. hemisphere.
The observed deviations are mainly caused by the geometrical limit of detection 
at $\Theta_{\rm lab}=30^{\circ}$.
This cut biases down the $F_{\rm S}^{(0)}$ parameter 
by about $20\%$ for $Z=1$ and by less than $10\%$ for heavier particles
in the PM5 event class at $E_{\rm lab}=250$$A$\,MeV.
Note that these effects decrease with increasing fragment size~\cite{cro 95}
because heavy particles, due to their low sensitivity to thermal fluctuations,
occupy a smaller phase space.
Therefore, in order to avoid misleading interpretations of the data, 
the $F_{\rm S}^{(0)}$ flow parameters presented in what follows
include only the measurements of particles whose charge is $\geq 2$.
On the other hand, it is worth noting that the effects of the 
$\Theta_{\rm lab}=30^{\circ}$
cut decreases with increasing impact parameters since peripheral 
and semi-central event topologies are less accentuated in the transverse 
direction as compared to central events.

\subsection{Squeeze-out} 

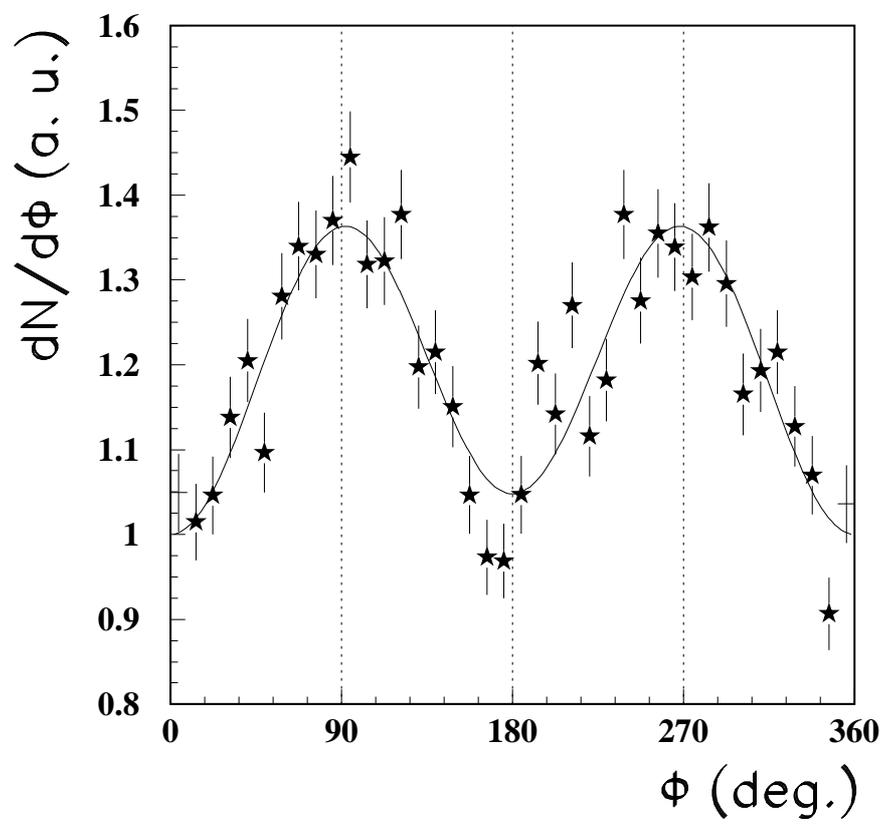
\begin{figure}
\caption{Azimuthal distribution of mid-rapidity
($-0.1 < y^{(0)} < 0.1$) $Z=3$ particles measured in semi-central
PM4 collisions at $E_{\rm lab}=250$$A$\,MeV.
$\Phi$ is the particle azimuthal angle around the beam axis, with respect to
the reaction plane.
The distribution is extracted with the transverse momentum cut
$0.4 < p_{\rm t}^{(0)} < 0.55$.
The solid curve is the result of the fit described in the text.
Error bars correspond to statistical uncertainties.}
\label{fig2}
\end{figure}

The out-of-plane flow component was investigated from the azimuthal 
distributions d$N$/d$\Phi$ ($\Phi$ is the azimuthal angle of the detected 
particle relative to the azimuth of the reaction plane)
around the beam axis of mid-rapidity particles,
selected by imposing a rapidity cutoff $-0.1<y^{(0)}<0.1$.
It is now an established fact that the out-of-plane 
anisotropy increases strongly with the transverse momentum of charged 
particles~\cite{bri 96,bas 96,lam 94}.
Therefore, in order to extract relevant information from the data, 
we determined the squeeze-out signal by choosing a $p_{\rm t}$ cut
which, within the acceptance, gives access to the largest momenta and 
sufficiently wide for statistics considerations.
This $p_{\rm t}$ window is $0.4 < p_{\rm t}^{(0)} < 0.55$, 
$p_{\rm t}^{(0)}$ being the particle transverse momentum per nucleon
divided by the projectile momentum per nucleon in the c.m. system.
These rapidity and transverse momentum cuts used to extract  
the signal, define a portion of the phase space
which is covered by the FOPI detector acceptance~\cite{bas 96}.
Figure~\ref{fig2} shows a typical d$N$/d$\Phi$ distribution for $Z=3$ particles 
in the PM4 event class at an incident energy of $E_{\rm lab}=250$$A$\,MeV.
A clear preferential emission is observed in the direction perpendicular
to the reaction plane ($\Phi=90^{\circ}$ and $\Phi=270^{\circ}$).
This enhanced emission reflects the squeeze-out effect.
The magnitude of the latter is commonly defined as the ratio $R_{\rm N}$ 
of the number
of particles emitted perpendicular to the reaction plane to the number
of particles emitted in the reaction plane~\cite{dem 90,gut 90} :
$R_{\rm N}=(N(90^{\circ})+N(270^{\circ}))/(N(0^{\circ})+N(180^{\circ})).$
The $R_{\rm N}$ ratio is extracted by fitting a function of the form 
$N(\Phi)=a_0+a_1\times\cos(\Phi)+a_2\times\cos(2\Phi)$ to the data 
(curve of Fig.~\ref{fig2}).
Thus $R_{\rm N}$ is calculated as $R_{\rm N}=(a_0 - a_2)/(a_0 + a_2)$.
According to this definition, $R_{\rm N}<1$ and $R_{\rm N}>1$ are related 
to a preferential
emission of matter in the reaction plane and out of this plane, respectively
while $R_{\rm N}=1$ corresponds to a perfect azimuthally isotropic situation.
The anisotropy ratio $R_{\rm N}$ can be corrected, like the flow parameter
$F_{\rm S}^{(0)}$, for the uncertainties due to fluctuations of the reaction 
plane~\cite{dem 90}.
However, we found with the help of simulations that,
for the low multiplicity events, the values of
$<\cos^2(\Delta\Phi_{\rm R})>$, which are the quantities
involved in these corrections, were not determined with good accuracy.
Thus, $R_{\rm N}$ ratios reported in what follows are not corrected for 
the effects
of reaction plane fluctuations.
Nevertheless, the possible influence of these effects on the observed trends
will be discussed in the following.

\subsection{Balance energy}

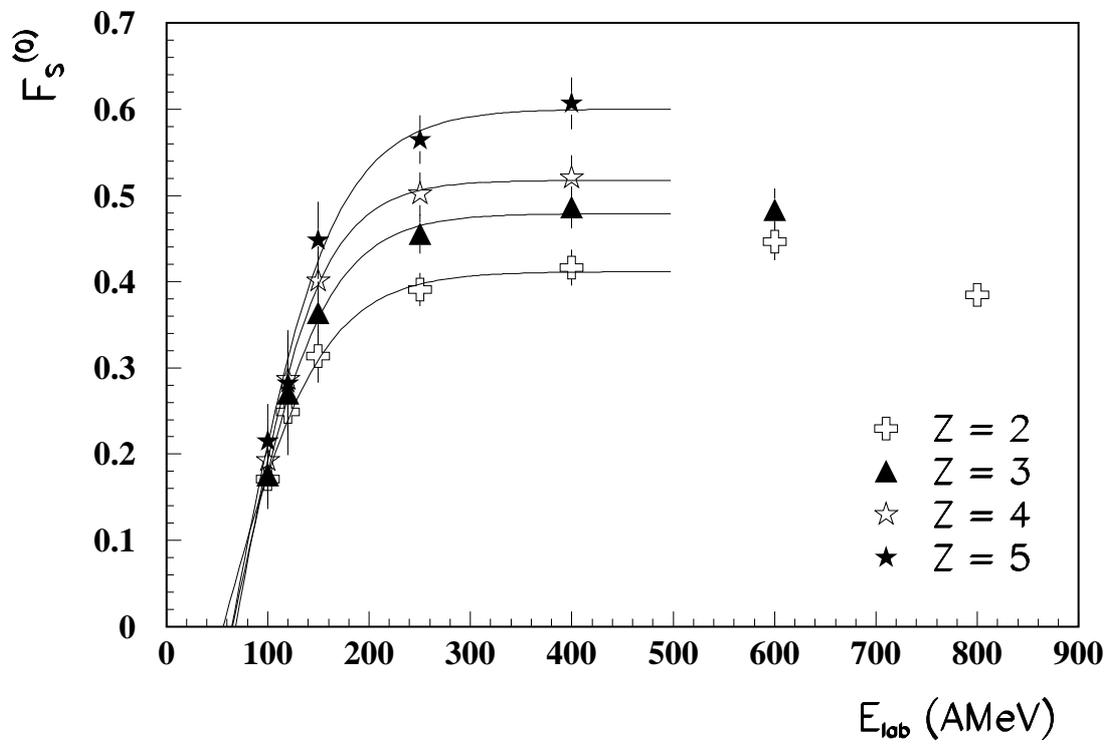
\begin{figure}
\caption{Excitation functions of the normalized
flow parameter
$F_{\rm S}^{(0)}$ measured for different particles ($Z=2$ to 5)
in semi-central PM4 collisions.
The values are corrected for fluctuations of the estimated reaction plane.
The solid lines correspond to the fits to the data points
from $E_{\rm lab}=100$ to 400$A$\,MeV with Fermi functions.
Error bars correspond to systematic uncertainties, estimated to
20, 20, and 10$\%$ for $E_{\rm lab}=100$, 120, and 150$A$\,MeV,
respectively and less than $10\%$ for $E_{\rm lab} \geq$ 250$A$\,MeV.
These errors are larger than statistical uncertainties.}
\label{fig3}
\end{figure}

Figure~\ref{fig3} shows the excitation functions of the scale invariant flow 
parameter $F_{\rm S}^{(0)}$ for different particles ($Z=$ 2 to 5).
A sudden decrease is observed in the incident energy region 
$E_{\rm lab} < 200$$A$MeV. It is interesting to notice that this 
sudden change is much more pronounced in the case of the heavier
fragments which are more sensitive to the collective motion. 
An extrapolation with Fermi functions
allows us to estimate the balance energy (intersection with the
abscissa) for different types of particles. 
Note that the balance energy values extracted from extrapolations of the
data with other functions, such as logarithmic and second order polynomial
ones, were found to be the same within error bars~\cite{cro 96a}. 
We have also verified that the use of other scaling variables for the in-plane
transverse momentum, such as $<p_{\rm x}/p_{\rm t}>$ 
(as used in~\cite{zha 90}) or $<p_{\rm x}>/<p_{\rm t}>$, 
leads to very similar results.
The extrapolation for $Z=3$
fragments leads to an intersection energy of 
$E_{\rm BAL} = 65\pm 15$$A$\,MeV 
for events corresponding to the PM4 bin.
The resulting values for the other particles ($Z=2$, 4 and 5) are the 
same within uncertainties as the one obtained for $Z=3$ (see Tab.~\ref{tab3}).
This confirms the observations established by studying lighter systems 
that the balance energy is independent of the size of the
detected particle~\cite{ogi 90,wes 93}.
Our present balance energy point is somewhat larger than the one 
obtained for the same system from other experiments~\cite{zha 90,par 95}.
This is probably due to the fact that our PM4 event class 
contains less central events than the one used in 
references~\cite{zha 90,par 95}. 
Indeed, as it is shown in the following, the balance energy 
is found to decrease with decreasing impact parameters.
With this in mind, the balance energy value extracted here is
consistent with the systematics of the balance energy as a function of
the mass of the combined projectile-target system obtained
from MSU and GANIL results~\cite{but 95b}.
On the other hand, it is worth noting that for a heavy system like Au+Au, 
because of the strong Coulomb repulsion, the overall force is always 
repulsive~\cite{sof 95}.
Since the balance energy should correspond to the energy at which 
the attractive and repulsive component of the nuclear interaction
balance each other, it must be evaluated without contamination of non nuclear 
contribution to the flow.
For light systems this problem is less severe because of the much weaker 
Coulomb repulsion.
Nevertheless, for heavy systems one may hope to extract 
the correct value of the balance energy by extrapolating the 
$F_{\rm S}^{(0)}$
values from sufficiently high energies where Coulomb contribution is
negligible.    

\subsection{Transition energy}

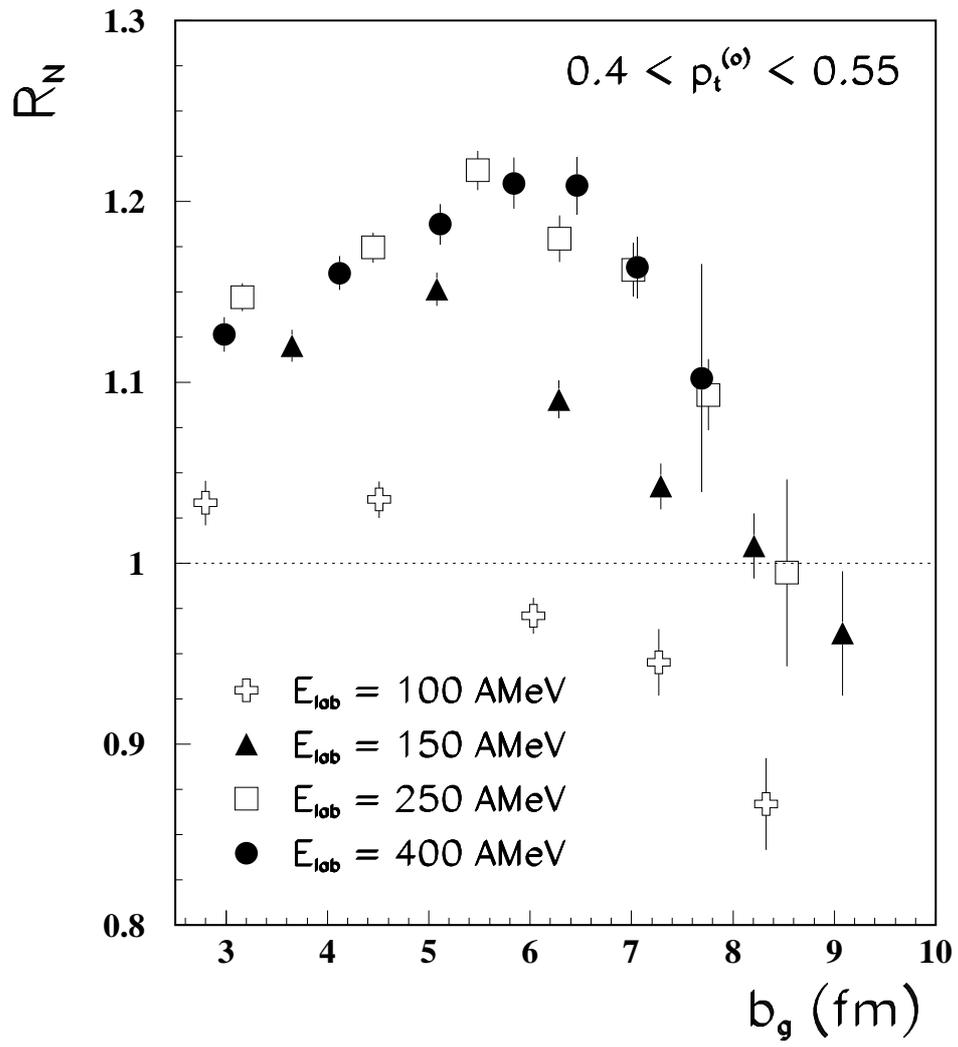
\begin{figure}
\caption{Anisotropy ratio $R_{\rm N}$ as a function of the
geometrical
impact parameter $b_{\rm g}$ for Au+Au collisions at $E_{\rm lab}=$100$A$\,MeV
(crosses),
150$A$\,MeV (triangles), 250$A$\,MeV (squares) and 400$A$\,MeV (circles).
$R_{\rm N}$ includes the contributions of all detected particles
with $0.4 < p_{\rm t}^{(0)} < 0.55$, each being weighted by its charge.
Error bars correspond to statistical uncertainties.}
\label{fig4}
\end{figure}

The dependence of the anisotropy ratio $R_{\rm N}$ on the collision impact 
parameter 
is presented in Fig.~\ref{fig4} at four incident energies going from 100 to 
400$A$\,MeV.
The signal includes here the contributions of all detected particles 
each weighted by its charge.
By doing so we reconstruct a coalescence invariant quantity
which makes meaningful the investigation of  
the anisotropy ratio as a function of the impact parameter and the beam energy.
The geometrical impact parameter $b_{\rm g}$ was obtained from the measured 
multiplicity distributions by assuming a sharp-cut-off approximation.
This allows us to perform direct comparisons of data measured at different 
bombarding energies. 
As can be seen in Fig.~\ref{fig4}, the correlation between the $R_{\rm N}$ ratio
and the impact parameter exhibits a very different trend as the incident 
energy decreases.
At the highest bombarding energy (400$A$\,MeV) one can observe a bell-shaped
distribution whose maximum is located at intermediate impact parameters
(close to $6fm$).
It is worth noting that the results obtained at higher energies 
are quite similar to those observed at $E_{\rm lab}=400$$A$\,MeV~\cite{cro 96b}.
With decreasing beam energies, the shape of the correlation evolves
gradually toward a different trend which is an evidence for a clear change 
in the emission pattern.
Thus at $E_{\rm lab}=100$$A$\,MeV, with decreasing impact parameter, 
one observes
a transition from a preferential in-plane emission ($R_{\rm N} < 1$) 
to the squeeze-out effect
characterized by an enhanced out-of-plane emission ($R_{\rm N} > 1$).

Before going to the interpretations of this behaviour, it must be pointed
out that two effects might influence the $R_{\rm N}$ ratio :
i) the dispersion of the reconstructed reaction plane with respect to the true
one, which tends to attenuate the magnitude of the signal and ii) 
the sideward-flow deflection which favors the emission of particles
in the reaction plane. 
In both cases the magnitude of the effect is impact parameter and beam energy
dependent.
Therefore, in order to eliminate possible ambiguities in the 
interpretation of the experimental observations in Fig.~\ref{fig4}, it was 
necessary to examine the respective influences of these effects on the
correlation between the anisotropy ratio and the impact parameter.
A further complete analysis of the data has allowed us to show 
that the bell-like shape of the distribution observed at high beam energies
is preserved after taking into account both mentioned effects~\cite{cro 96b}.
On the other hand, the transition from $R_{\rm N}>1$ to $R_{\rm N}<1$ at 
$E_{\rm lab}=100$$A$\,MeV cannot be caused by one of these two effects.
Indeed, fluctuations of the reaction plane tend to attenuate an anisotropy 
signal
regardless of whether the $R_{\rm N}$ ratio is larger or smaller than 1.
Therefore, taking into account the corresponding corrections in Fig.~\ref{fig4},
the transition effect would be even more pronounced.
On the other hand, an extraction of the $R_{\rm N}$ quantity around the flow 
axis would shift up the experimental points but at large $b_{\rm g}$'s, where
the in-plane enhancement is observed, the flow angle is expected to be fairly 
low~\cite{cro 96a} in particular at the lowest beam energy 
($E_{\rm lab}=100$$A$\,MeV) which is close to the balance 
energy (Fig.~\ref{fig3}).

Let us now go back to the interpretations of the experimental observations
of Fig.~\ref{fig4}.
At high energies, the maximum located near $b_{\rm g} = 6fm$ is consistent with 
an expansion-shadowing picture,
{\em i.e.} an expansion of the compressed matter in the central region of the
collision which is hindered by the presence of cold spectator remnants.
It is worth noting that recent IQMD calculations for neutrons~\cite{bas 94}
predict a similar bell-shaped correlation with a maximum around $7 fm$.
At low incident energies ($E_{\rm lab}\le$ 150$A$\,MeV), a clear
evidence for a transition from an enhanced in-plane emission pattern
to a preferential out-of-plane emission is observed when $b_{\rm g}$ decreases.
The results show that this transition takes place close to
$E_{\rm lab} = 100$$A$\,MeV for collisions with impact parameters
$b_{\rm g} \simeq 6 fm$. 
This effect was already observed for a lighter system~\cite{pop 94}
and very recently for the same system Au+Au~\cite{tsa 96}.
It might be attributed to a change from a collective 
rotational behaviour governed by the attractive mean field at low energies,
to the high energy squeeze-out effect resulting from the repulsive pressure
built up during the high density stage of the collision~\cite{pop 94,tsa 96}.

\vspace{0.5cm}

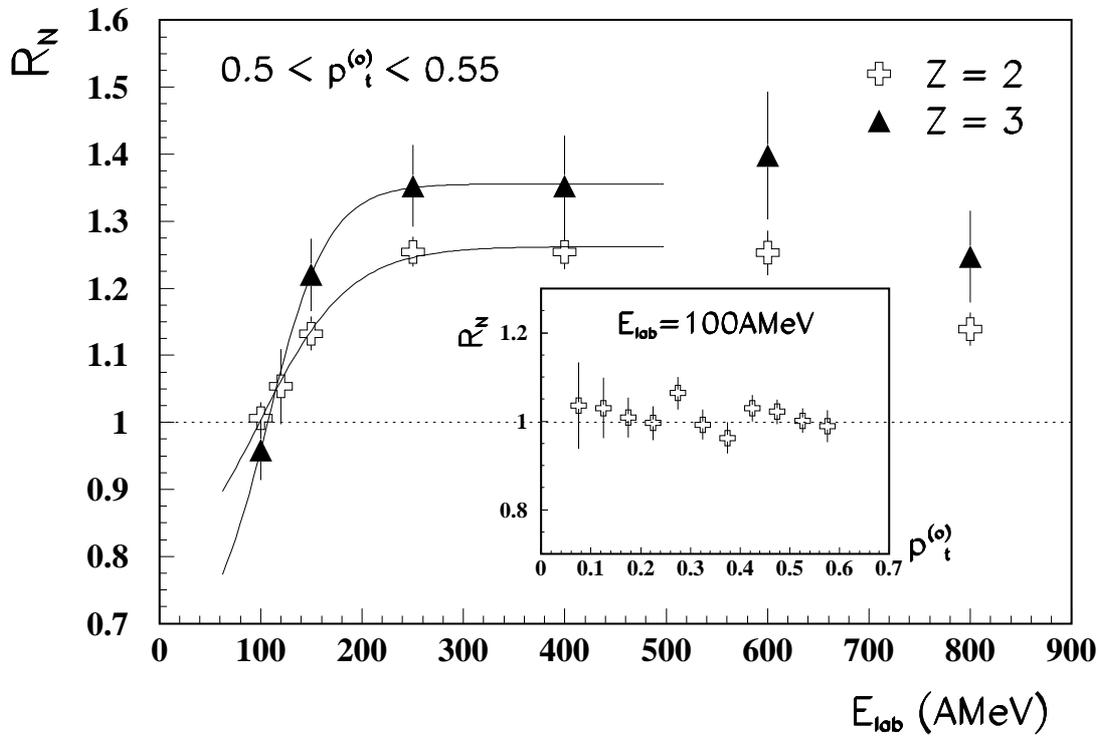
\begin{figure}
\caption{Excitation functions of the $R_{\rm N}$ ratio
for $Z=2$ (crosses) and $Z=3$ particles (triangles)
detected in semi-central PM4 collisions under the condition
$0.5 < p_{\rm t}^{(0)} < 0.55$.
The curves represent the results of a fit with a Fermi
function to the data points from $E_{\rm lab}=100$ to 400$A$\,MeV.
The insert shows the $p_{\rm t}^{(0)}$ dependence of the $R_{\rm N}$
ratio at an incident energy of 100$A$\,MeV.
Error bars correspond to statistical uncertainties.}
\label{fig5}
\end{figure}

The transition energy $E_{\rm TRA}$, corresponding to an azimuthally symmetric
distribution ($R_{\rm N} = 1$), can be evaluated from the
excitation function of the anisotropy ratio $R_{\rm N}$.
This has been already investigated in a previous analysis~\cite{bas 96} 
where the experimental $E_{\rm TRA}$ values were compared to the predictions
of the IQMD model.
The excitation functions of the $R_{\rm N}$ ratio are presented in 
Fig.~\ref{fig5} for $Z=2$ and 3 particles emitted in semi-central (PM4) 
collisions.
Since events were selected here over a large multiplicity bin, 
the large statistics allowed us to investigate the anisotropy signal 
in a reduced high $p_{\rm t}$ window as compared to the previous one.
This $p_{\rm t}$ condition (fixed as $0.5 < p_{\rm t}^{(0)} < 0.55$) 
was chosen in 
order to extract from the data the largest $R_{\rm N}$ magnitudes within 
the acceptance of the detector.  
As one can see from Fig.~\ref{fig5}, the $R_{\rm N}$ ratio tends to 
saturate above $E_{\rm lab} = 250$$A$\,MeV and seems possibly to decrease
at higher energies.
On the other hand, the behaviour of the signal at low beam energies
exhibits the same sudden change as in the case of the excitation 
functions of the sideward flow.

In order to extract the transition energy, the data points in the 
beam energy range 100-400$A$\,MeV were fitted with 
a Fermi function (curves of Fig.~\ref{fig5}). 
The values of $E_{\rm TRA}$, reported in Tab.~\ref{tab4}, 
were determined for $Z=2$ and 3 
particles at the intercept of $R_{\rm N}=1$ with the curves.
Note that the intersection energies were found to be quite insensitive
to the form of the fitting function.
They seem to be, as observed in the case of the balance energy, independent of
the type of the detected particle within the error bars.
It can be seen from the results presented in Tab.~\ref{tab3} and 
Tab.~\ref{tab4} that
the transition energy for the PM4 event class ($E_{\rm TRA} \simeq 100$$A$\,MeV)
is somewhat larger than the corresponding balance energy 
($E_{\rm BAL} \simeq 65$$A$\,MeV).
This finding agrees with recent theoretical calculations for the Ca+Ca
system~\cite{sof 95}. It could be explained by angular momentum effects 
or by the fact that
at the balance point the compression is not high enough to generate
an enhanced out-of-plane emission.

On the other hand, the $R_{\rm N}$ ratio, which increases 
with the particle transverse momentum for 
$E_{\rm lab} \geq 150$$A$\,MeV~\cite{bri 96,bas 96,lam 94}, 
has been extracted over a narrow 
$p_{\rm t}$ window ($0.5 < p^{(0)}_{\rm t} < 0.55$).
This could suggest that the extracted transition energy is $p_{\rm t}$ 
dependent.
Nevertheless, as shown by the insert in Fig.~\ref{fig5},
at $E_{\rm lab}=100$$A$\,MeV where the transition effect takes place, 
the $R_{\rm N}$ ratio remains nearly close to 1 in the whole explored 
$p^{(0)}_{\rm t}$ domain.
Finally, as mentioned before, the $R_{\rm N}$ ratios are not 
corrected for fluctuations of the reconstructed reaction plane because
the $<\cos^2(\Delta\Phi_{\rm R})>$ values cannot be accurately determined. 
Therefore, in order to estimate how the transition energy could be 
influenced by these 
fluctuations, we have extracted $E_{\rm TRA}$ from the $R_{\rm N}$ values
obtained by correcting the measured anisotropy ratios using 
the $<\cos^2(\Delta\Phi_{\rm R})>$ factors calculated for filtered 
IQMD theoretical events, where the true reaction plane is known.
The resulting transition energy value, for $Z=3$ particles, was found to be 
$E_{\rm TRA} = 111\pm 10$$A$MeV in the PM4 event class.
This value is very close to the one obtained without correction 
(Tab.~\ref{tab4}).   

\subsection{Centrality dependence of threshold energies}

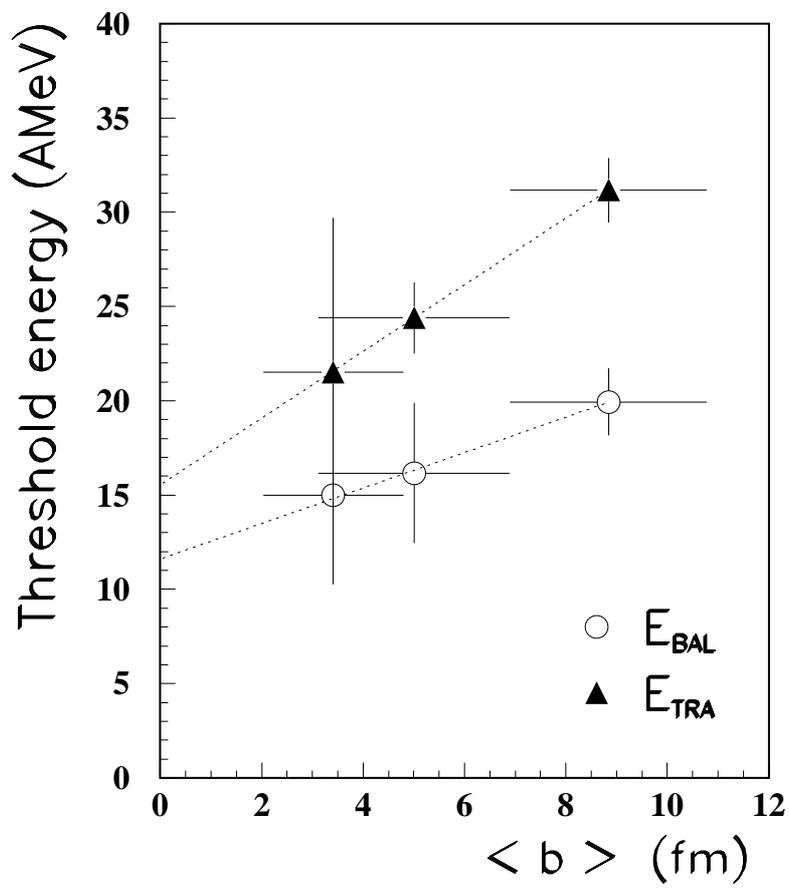
\begin{figure}
\caption{Impact parameter dependence of the c.m. threshold
energies $E_{\rm BAL}$ (circles) and $E_{\rm TRA}$ (triangles),
determined from the measurements of $Z=3$ and $Z=2$ particles, respectively.
$<b>$ has been determined within simulations using the IQMD model
at $E_{\rm lab}=150$$A$\,MeV (see Tab.2).
The dotted lines are linear fits to the data.
Horizontal error bars represent the r.m.s. deviation of $b$ distributions.
Vertical error bars correspond to systematic and statistical
errors for $E_{\rm BAL}$ and $E_{\rm TRA}$, respectively.}
\label{fig6}
\end{figure}

The balance and transition energies extracted from extrapolations
of the measured $F_{\rm S}^{(0)}$ and $R_{\rm N}$ excitation functions 
in the three different PM multiplicity bins are shown in Fig.~\ref{fig6} 
as a function
of the collision centrality.
$E_{\rm BAL}$ and $E_{\rm TRA}$ are 
expressed here not in terms of the projectile energy but of the corresponding
energy in the c.m. system.
They were obtained,
due to considerations of detector acceptance effects and statistics,
from the measured excitation functions of $Z=3$ for $E_{\rm BAL}$
and $Z=2$ for $E_{\rm TRA}$.
As it can be seen, the transition energy is larger than the balance energy 
over the whole explored impact parameter range.
Both threshold energies $E_{\rm BAL}$ and $E_{\rm TRA}$ increase with 
increasing impact parameters.
Such a behaviour has been also very recently observed in
the case of the balance energy of the Ar+Sc system~\cite{pak 96}
and is predicted by the IQMD model for the Ca+Ca system~\cite{sof 95}.
This increase of $E_{\rm BAL}$ and $E_{\rm TRA}$ with $b$ 
may indicate that the threshold of flow effects is related
to the energy deposited locally into the overlap zone of the collision :
if the local temperature is large enough, the generated pressure can 
overcome the attractive nuclear forces.
When going from peripheral to central collisions, the
deposited energy becomes larger and the onset of flow is therefore
expected to take place at a lower incident energy.
At bombarding energies close to the Fermi energy, the participant picture 
is not well developed like at higher energies and the local temperature 
depends sensitively on the local heat relaxation-time and on collision time.
With this in mind, it is interesting to compare the threshold
energies in finite impact parameter collisions with the threshold energy
for the radial expansion in central collisions.
The latter was determined in reference~\cite{gob 95} by  
extrapolating the excitation function of the mean radial
flow velocity, measured in central collisions, toward low incident energies.
Its value was found to be about 
$E_{\rm RAD}=35 \pm 10$$A$\,MeV~\cite{gob 95} which correponds to a 
c.m. energy of $8.7 \pm 2.5$$A$\,MeV. 
Note that this value is consistent with recent results for the Au+Au reaction
obtained at $E_{\rm lab}=35$$A$\,MeV~\cite{ago 96}.
Now, considering again Fig.~\ref{fig6}, 
a rough linear extrapolation of our balance and transition energy points 
toward $<b>$ = 0 leads to values of $12\pm 6$$A$\,MeV 
and $16 \pm 5$$A$\,MeV, respectively.
These values are close to each other and consistent with the value of 
$E_{\rm RAD}$ within the error bars.
It is still premature to go to more quantitative interpretations
because of the large experimental uncertainties. 
Nevertheless, it is tempting to speculate on the basis of the observations
in Fig.~\ref{fig6} on a common phenomenon at the origin of radial flow, 
sideward-flow and squeeze-out
effects which takes place for the same deposited energy : when nuclear
matter is heated above a certain limit, the attractive nuclear forces
are counterbalanced by the thermal pressure and the system starts to expand
above this threshold at which the overall force becomes repulsive.
This expansion of nuclear matter tends to an azimuthally symmetric 
pattern (radial flow) in highly central collisions,  
while in semi-central collisions the expanding matter is pushed 
to the side (sideward-flow) and 
the presence of cold spectator remnants hinders participant nucleons
to escape in the reaction plane, which gives rise to a preferential 
out-of-plane emission (squeeze-out). 
This scenario for the squeeze-out effect is
consistent with the observed correlation between  
the anisotropy ratio and the impact parameter (Fig.~\ref{fig4}).
Differences on the various threshold energies can be expected on the bases
of effects like collision geometry, system size, Coulomb contribution, 
angular momentum dissipation, non equilibrium dynamics, {\em etc}.
On the other hand, it is worth noting that the c.m. energy of 8.7$A$\,MeV
where the radial flow sets in,
is close to the one of the caloric curve, reported by the ALADIN
Collaboration~\cite{poc 95,pap 95}, where the temperature grows strongly 
again after the presently debated plateau. 
As pointed out in ref.~\cite{bon 87}, this observation suggests that the onset 
of flow could be interpreted as a possible signature of a liquid-gas phase 
transition.
Indeed, as discussed before, the onset of flow phenomena indicates 
a change in the reaction scenario from a global repulsive mechanism 
to a global attractive mechanism when decreasing the incident energy.
In a fluid dynamical vision of the collision, it could also be seen as 
a manifestation of a liquid-gas phase transition since this leads to a 
sudden decrease of the repulsive pressure and, consequently, of the flow 
magnitude.    

\section{Conclusion}

Collective effects of nuclear matter in Au+Au collisions at incident
energies ranging from 100$A$\,MeV to 800$A$\,MeV were measured for 
light particles 
and intermediate mass fragments with the FOPI detector at GSI.
The centrality and beam energy dependence of both sideward-flow and squeeze-out
effects were investigated.
The evolution of the squeeze-out magnitude $R_{\rm N}$ with the impact parameter
is found to change drastically with the incident energy.
At high incident energies the squeeze-out signal exhibits a bell-shape 
with a maximum located at intermediate impact parameters.
This trend is consistent with an expansion-shadowing scenario where
the expansion of highly compressed participant nuclear matter is hindered 
by the presence of cold spectator remnants.
At low energies, the data show clearly a transition from an in-plane
preferential emission to an out-of-plane enhancement when the centrality 
increases.
This phenomenon might be attributed to a change from a 
collective rotational
motion for large impact parameters to the squeeze-out effect for smaller 
impact parameters~\cite{pop 94,tsa 96}.
The balance and transition energies, corresponding to the onset of
sideward-flow and squeeze-out effects, respectively, have been evaluated 
from extrapolations toward lower beam energies of the excitation 
functions of the scale invariant flow parameter $F_{\rm S}^{(0)}$
and the anisotropy ratio $R_{\rm N}$.
Both of them are found to be, within errors, independent of the size of the
detected particle.
The transition energy is larger than the balance energy.
The extrapolation of the centrality dependence of both threshold energies 
toward $b = 0$ leads to a value which is close to the threshold energy for
the radial expansion in central collisions.
This suggests that the same phenomenon could be at the origin of the three 
processes.
The relevant parameter seems to be the energy deposited into the system
in order to counterbalance the attractive nuclear forces although 
compression effects can also be present.
In central collisions the repulsive pressure is expected to be the highest 
and the participant matter can expand freely in all directions, while 
for $b \neq 0$, besides complex geometrical and dynamical effects, the presence 
of the spectator matter causes the appearence of sideward-flow and squeeze-out.
Comparisons of the present experimental results
with transport model predictions should provide interesting information
about the in-medium reduction of the nucleon-nucleon cross 
section~\cite{xu 91,zho 93,tsa 96}.

\section*{Acknowledgement}

This work was supported in part by the French-German agreement between
GSI and IN2P3/CEA and by the PROCOPE-Program of the DAAD.


\begin{table}
\caption{Lower limit PM5$_{\;\mbox{l}}$ of the PM5
multiplicity bin at the different beam energies.}
\vspace{0.5cm}
\begin{tabular}{cccccccc} 
$E_{\rm lab}$ ($A$\,MeV) & 100 & 120 & 150 & 250 & 400 & 600 & 800 \\ \hline
PM5$_{\;\mbox{l}}$   & 28 & 31 & 36 & 44 & 55 & 62 & 70 \\
\end{tabular}
\label{tab1}
\end{table}

\vspace{1.cm}

\begin{table}
\caption{Mean impact parameter $<b>$ for each experimental
PM multiplicity bin at incident energies of 150 and 400$A$\,MeV.
$<b>$ is determined in the framework of the IQMD model
with a hard momentum dependent interaction.
Errors represent the r.m.s. deviation of the $b$ distributions.}
\vspace{0.5cm}
\begin{tabular}{cccc} 
Multiplicity bin & PM3 & PM4 & PM5  \\ \hline
$E_{\rm lab}$=150$A$\,MeV
&$8.8\pm 1.9$&$5.0\pm 1.9$&$3.4\pm 1.4$ \\
$E_{\rm lab}$=400$A$\,MeV
&$7.0\pm 1.1$&$4.1\pm 1.5$&$3.2\pm 1.1$ \\
\end{tabular}
\label{tab2}
\end{table}

\vspace{1.cm}

\begin{table}
\caption{Balance energy for $Z=2$ to 5 particles
under the PM4 multiplicity cut (see text). Errors correspond to
systematic uncertainties.}
\vspace{0.5cm}
\begin{tabular}{ccccc} 
$Z$ & 2 & 3 & 4 & 5  \\ \hline
$ E_{\rm BAL}$ ($A$\,MeV) &$56.0\pm 21.4$&$65.2\pm 14.9$&$
68.3\pm 11.1$&$64.5\pm 15.1$ \\
\end{tabular}
\label{tab3}
\end{table}

\vspace{1.cm}

\begin{table}
\caption{Transition energy for $Z=2$ and 3 particles
under the PM4 multiplicity cut (see text). Errors correspond to statistical
uncertainties only.}
\vspace{0.5cm}
\begin{tabular}{ccc} 
$Z$ & 2  & 3 \\ \hline
$ E_{\rm TRA}$ ($A$\,MeV) &$98.9\pm 7.6$&$107.0\pm 6.9$\\
\end{tabular}
\label{tab4}
\end{table}


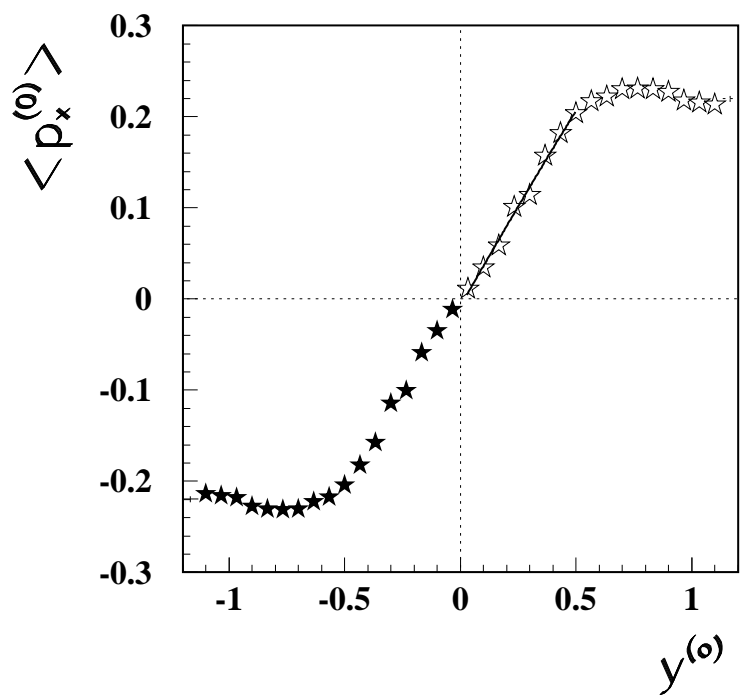
\end{document}